\documentclass[preprint,11pt]{aastex}

\shortauthors{}
\shorttitle{}

\begin{document}

%
\def\ltsima{$\; \buildrel < \over \sim \;$}
\def\lsim{\lower.5ex\hbox{\ltsima}}
\def\gtsima{$\; \buildrel > \over \sim \;$}
\def\gsim{\lower.5ex\hbox{\gtsima}}
                                                                                          
%

\bibliographystyle{apj}

\title{Precise Radius Estimates for the Exoplanets WASP-1b and WASP-2b}

\author{
David Charbonneau\altaffilmark{1,2},
Joshua N. Winn\altaffilmark{3},
Mark E. Everett\altaffilmark{4},
David W. Latham\altaffilmark{1},
Matthew J. Holman\altaffilmark{1},
Gilbert A. Esquerdo\altaffilmark{1,4},
and Francis T. O'Donovan\altaffilmark{5}
}

\altaffiltext{1}{Harvard-Smithsonian Center for Astrophysics,
  60 Garden Street, Cambridge, MA 02138; dcharbonneau@cfa.harvard.edu.}
\altaffiltext{2}{Alfred P. Sloan Research Fellow}
\altaffiltext{3}{Department of Physics, and Kavli Institute for
  Astrophysics and Space Research, Massachusetts Institute of
  Technology, Cambridge, MA 02139.}
\altaffiltext{4}{Planetary Science Institute, 1700 East Fort Lowell Road,
  Suite 106, Tucson, AZ 85719.}
\altaffiltext{5}{California Institute of Technology, 
  1200 East California Boulevard, Pasadena, CA 91107.}

\begin{abstract}
We present precise $z$-band photometric time series spanning times of
transit of the two exoplanets recently discovered by the SuperWASP
collaboration. We find planetary radii of $1.44 \pm 0.08\, R_{\rm J}$
and $1.04 \pm 0.06\, R_{\rm J}$ for WASP-1b and WASP-2b, respectively.
These error estimates include both random errors in the photometry and
also the uncertainty in the stellar masses. Our results are 5 times
more precise than the values derived from the discovery data alone.
Our measurement of the radius of WASP-2b agrees with previously
published models of hot Jupiters that include both a 20-$M_{\Earth}$
core of solid material and the effects of stellar insolation. In
contrast, we find that the models cannot account for the large size of
WASP-1b, even if the planet has no core. Thus, we add WASP-1b to
the growing list of hot Jupiters that are larger than expected. This
suggests that ``inflated'' hot Jupiters are more common than
previously thought, and that any purported explanations involving
highly unusual circumstances are disfavored.
\end{abstract}


\section{Introduction}

The wide-field surveys for transiting exoplanets have finally begun to
strike gold. For nearly 10 years, numerous groups have attempted to
use small-aperture lenses to identify transits of bright stars over
large patches of the sky. This turned out to be much more difficult
than initially expected, and the first success was achieved only two
years ago (Alonso et al.~2004). Since then, progress has accelerated,
and in the month of September 2006 alone, three different survey teams
announced the discovery of four transiting exoplanets.

The Trans-atlantic Exoplanet Survey (TrES) reported the discovery of
their second planet, TrES-2 (O'Donovan et al.\ 2006), the first
extrasolar planet detected in the field of view of the NASA {\it
Kepler} mission (Borucki et al.\ 2003) and the most massive exoplanet
known to transit a nearby bright star. The HATNet project announced
the discovery of HAT-P-1b (Bakos et al.\ 2006a), a hot Jupiter
orbiting one star of a visual binary, and the lowest-density hot
Jupiter yet detected. Most recently, the SuperWASP team announced the
discovery of two planets, WASP-1b and WASP-2b (Collier Cameron et al.\
2006), that are the subject of this paper. Thus, including the
discovery of a planet by the XO project earlier this year (McCullough
et al.\ 2006), four independent teams have now demonstrated the
capability to identify transiting hot Jupiters using very modest
(typically 10~cm) aperture automated observatories. Several more
projects also seem poised for success (for a review of current and
near-future transit-search projects, see Charbonneau et al.\ 2006a).

The reason why transiting planets are so precious, and why the
exoplanet community is ebullient over the progress in finding them, is
that only for transiting planets can one measure both the mass and the
radius. This in turn permits one to confront observations with
theoretical models of planetary structure. For the moment, this
confrontation is limited to the interesting case of the hot Jupiters,
for the simple reason that close-in planets are much more likely to
exhibit transits.

Prior to the detection of such objects in transiting configurations,
our naive expectation was that hot Jupiters would be similar to
Jupiter in structure, with a modest increase in radius due to the
effects of stellar insolation (e.g. Guillot et al.\ 1996; Lin, 
Bodenheimer, \& Richardson 1996).  However,
among the 14 cases that have since been discovered, there is a large
range in measured radii. At one extreme lies HD~149026b (Sato et al.\
2005; Charbonneau et al.\ 2006b), whose small radius bespeaks a
central core of solid material that composes roughly 70\% of the
planet by mass.  At the other extreme is HD~209458b (Knutson et al.\
2006), whose radius significantly exceeds the predictions of insolated
structural models (see, e.g., Baraffe et al.~2003 or Bodenheimer et
al.\ 2003, and for a contrary view, Burrows et al.~2003). The recently
discovered planet HAT-P-1b (Bakos et al.\ 2006a) is also larger than
theoretically expected. This suggests that in those two planets, at
least, there is an source of internal heat that was overlooked by
theoreticians.

Various mechanisms for producing this heat have been explored,
although certainly not exhaustively.  Bodenheimer et al.\ (2001; 2003)
proposed that there is a third body in the system that excites the
eccentricity of the hot Jupiter.  Ongoing tidal dissipation would then
provide the requisite energy, even if the mean eccentricity were as
small as a few per cent. However, subsequent observations have placed
an upper bound on the current eccentricity below the value required (Deming
et al.~2005, Laughlin et al.~2005a, Winn et al.~2005), and they have
not revealed any third body. Showman \& Guillot (2002) argued instead
that the heat could be provided by the conversion of several per cent
of the incident stellar radiation into mechanical energy that is
subsequently transported deep into the planetary
interior. Alternatively, Winn \& Holman (2005) invoked ongoing tidal
dissipation due to a nonzero planetary obliquity. Ordinarily, the
obliquity would be driven to very small values, but it is possible for
hot Jupiters to exist in a stable Cassini state (a resonance between
spin and orbital precession) with a significant obliquity.

Although measurements of either the winds or the spin states of hot
Jupiters may not be forthcoming soon, a possible avenue for progress
would be to examine the rate of occurrence of the anomalously-large
hot Jupiters relative to the hot Jupiter population as a whole (being
mindful of the observational biases that favor the detection of large
planets, as quantified by Gaudi 2005). In particular, the most
puzzling aspect of the Showman \& Guillot (2002) mechanism is why it
should act on some but not all hot Jupiters. Conversely, the Cassini
state described by Winn \& Holman (2005) requires some fine tuning,
making it an unattractive explanation if ``inflated'' planets turn out
to be relatively common.

Although the detection of the planets WASP-1b and WASP-2b (Collier
Cameron et al.\ 2006) is an important opportunity to address these
questions, the range of allowable planetary radii, $1.33 < R_{p} /
R_{\rm J} < 2.53$ for WASP-1b and $0.65 < R_{p} / R_{\rm J} < 1.26$
for WASP-2b, is too broad to meaningfully constrain the models.  In
this paper, we present the analysis of newly-acquired photometric time
series that serve to reduce the uncertainties in the radii of both
planets by a factor of 5.  We then interpret the new radius estimates
in the context of the known hot Jupiters and the published models of
their physical structural models.  We end by noting particular
opportunities for further follow-up presented by both planets.

\section{Observations}

We observed WASP-1 and 2 on the nights of predicted transits, with the
1.2~m telescope of the Fred L.\ Whipple Observatory on Mt. Hopkins,
Arizona. The WASP-1 transit occurred on UT~2006~September~27, while
the WASP-2 transit was on UT~2006~September~30. On each night, we used
Keplercam to obtain a continuous sequence of 30~s integrations of the
target and surrounding field stars. We employed the SDSS $z$ filter,
the reddest band available, to minimize the effects of differential
extinction on the photometry and the effect of stellar limb darkening on the
light curve. Keplercam employs a single 4096$\times$4096 Fairchild 486
CCD; we used 2$\times$2 binning. With a readout time of 9~s and total
overhead of 12~s between exposures, the resulting cadence was 42~s.
The field-of-view is $23\arcmin\times23\arcmin$ with a
0.67$\arcsec$~pixel$^{-1}$ plate scale when binned.  We used the
offset guider to maintain the telescope pointing to within $5\arcsec$
throughout the night. On each night, we started observing well before
the predicted time of ingress and ended well after egress.

For the WASP-1 event, we gathered 832 images over a timespan of 9.7
hours, spanning an airmass range of 1.0 to 2.1 that reached its
minimum value in the middle of the observing sequence. Light clouds
were present during the first hour, and conditions were photometric
afterwards. Since the hour in which clouds were present occurred well
before ingress, we decided to exclude those data in the analysis. The
full width at half-maximum (FWHM) of the stellar images was typically
$1\farcs6$, but occasionally degraded to $4\arcsec$.  For the WASP-2
event, we gathered 426 images spanning a period of 4.9 hours under
clear skies and spanning an airmass that began at 1.1 and increased to
2.1 over the observing sequence.  The seeing was stable, varying only
modestly between $1\farcs5$ and $1\farcs9$.  For calibration purposes,
we obtained on both nights dome flats and twilight sky flats along
with a set of bias images.

\section{Data Reduction}

To calibrate the images, we first subtracted an amplifier-dependent
overscan bias level and then joined the images from each quadrant into
a single frame.  We filtered the bias images from each night of
deviant pixels and averaged the cleaned biases to produce an average
bias frame.  We then used these average bias frames to subtract a
residual spatially-dependent bias pattern from the science images.  We
scaled our sky flat images to the same mean flux, and then averaged
them (while filtering out deviant pixels) to produce nightly
flat-field images, which we then use to flat-field each science image.

We performed aperture photometry using the IRAF\footnote{IRAF is
  distributed by the National Optical Astronomy Observatories, which
  are operated by the Association of Universities for Research in
  Astronomy, Inc., under cooperative agreement with the National
  Science Foundation.} PHOT task, which yielded
estimates of the instrumental magnitudes and sky magnitudes for the
target and comparison stars.  We estimated the sky magnitudes from the
median value in an annulus centered on the star after iteratively
rejecting pixel values that deviated by more than 3 standard
deviations from the mean.  To obtain differential photometry of the
target, we selected a group of field stars that were isolated and
located on a portion of the detector that was cosmetically clean.  We
then calculated the statistically-weighted mean magnitude of the
comparison stars in each frame as follows: We estimated the
photometric uncertainties based on the expectations of photon noise
from both the star and underlying sky, as well as detector read noise
and scintillation (as formulated by Gilliland et al.\ 1993).  We then
subtracted the mean magnitude of the comparison stars from those of
all stars in our list, including the target star.  We eliminated from
the list any comparison star found to be variable or exhibiting a
systematic trend in its brightness over time.  We iteratively
re-calculated the differential correction in this manner, eliminating
suspect comparison stars until we visually confirmed in plots of the
light curves that all of the comparison stars did not vary outside of
the expectations of the noise sources listed above.  This procedure
yielded 9 comparison stars for the WASP-1 data and 6 comparison stars
for the WASP-2 data.  We selected the optimal photometric aperture
(which depends primarily on the typical nightly seeing) and sky
annulus to be the ones that minimized the RMS deviation of the
out-of-transit portions of the differential light curve of the target
star.  We selected photometric apertures with radii of $6\farcs4$ and
$5\farcs4$ for the WASP-1 and WASP-2 data, respectively.  For both
nights, we selected an aperture for the sky annulus that spanned
$8\arcsec$ to $21\arcsec$.

Although the relative photometry removes the first-order effects of
extinction, color-dependent effects are not removed.  Stars of
different colors are extinguished by different amounts through a given
airmass.  For this reason, we applied a residual extinction correction
to the data.  The correction function was determined as part of the
model-fitting procedure that we describe in \S4.

\begin{figure}[p]
\epsscale{1.0}
\plotone{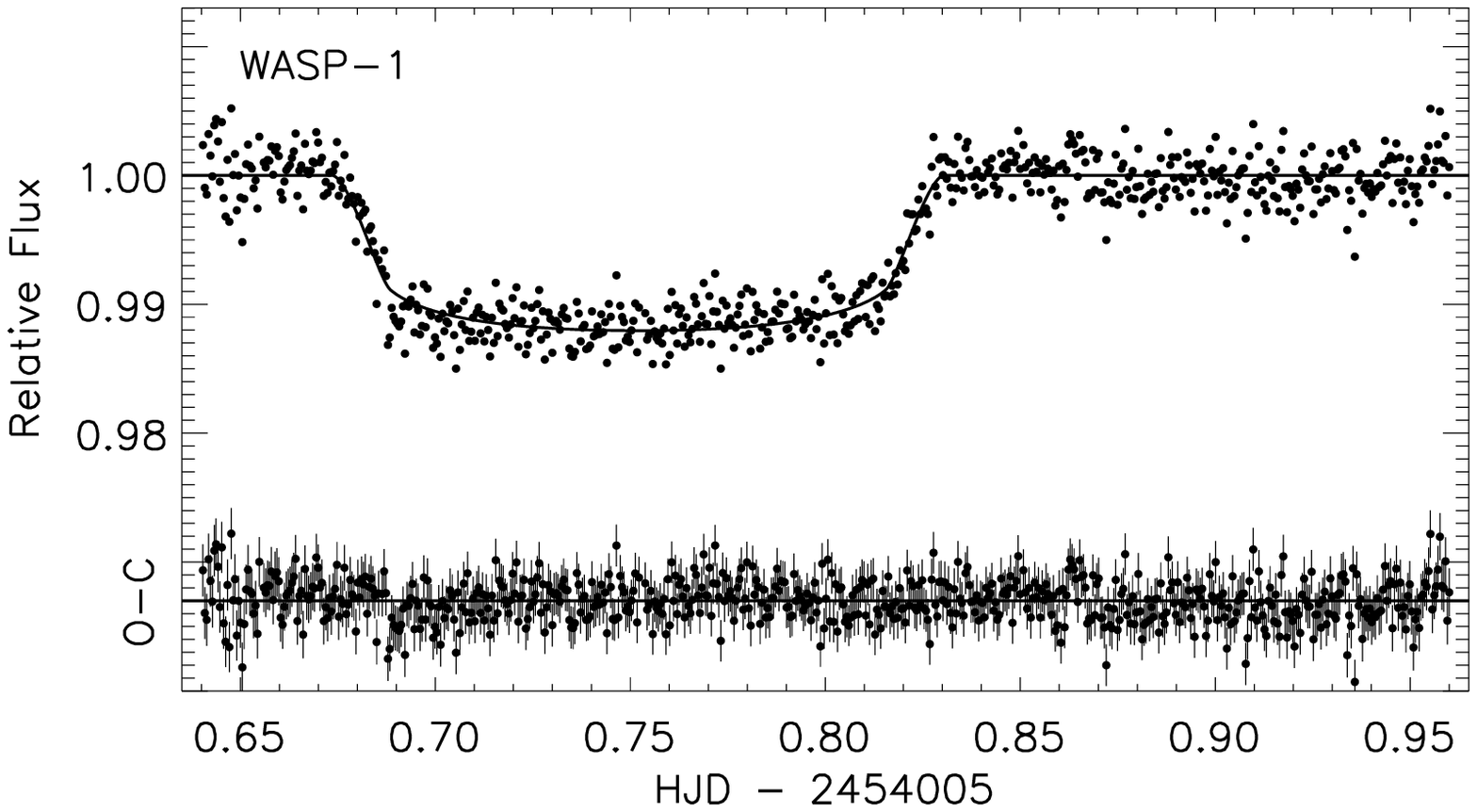}
\plotone{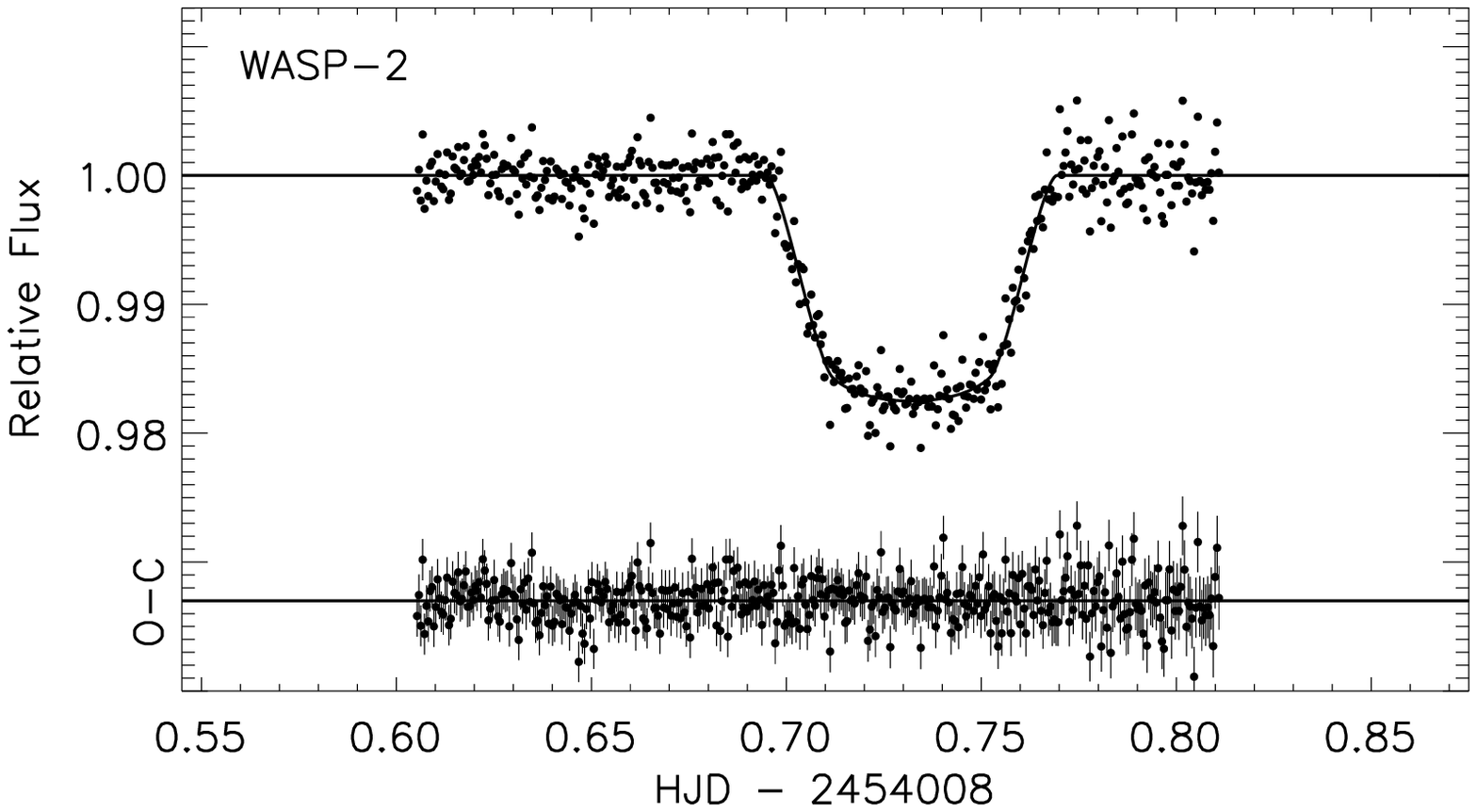}
\caption{
Relative $z$ band photometry of WASP-1 and WASP-2. The best-fitting
model is shown as a solid line. The residuals
(observed~$-$~calculated) and the rescaled $1~\sigma$ error bars are
also shown. The residuals have zero mean but are offset for
clarity by a constant flux so as to appear beneath each light curve. 
For both time series, the median time between exposures is 42~s,
and the RMS residual is 0.17\%.  The span of the axes is the same
in both plots, permitting a visual comparison of both events.  
The WASP-1b transit is longer and shallower,
as it corresponds to a more equatorial transit of a larger star.
\label{fig:lc}}
\end{figure}

The final photometry is given in Tables~1 and 2, and is plotted in
Fig.~\ref{fig:lc}. The fluxes and their uncertainties reported in the
tables have already been corrected for extinction.  The reported
uncertainties have been further rescaled such that $\chi^2/N_{\rm DOF}
= 1$ for the best-fitting model. The scaling factors were determined
independently for each night, but turned out to be nearly the same:
1.28 for the WASP-1 data and 1.29 for the WASP-2 data.

\section{The Model}

We used the same modeling techniques that have been employed
previously by the Transit Light Curve (TLC) project (e.g. Holman et
al.~2006a; Winn et al.~2006). Our model is based on a star (with mass
$M_{\star}$ and radius $R_{\star}$) and a planet (with mass $M_p$ and
radius $R_p$) in a circular orbit with period $P$ and inclination $i$
relative to the plane of the sky. We define the coordinate system such
that $0\arcdeg \leq i\leq 90\arcdeg$. We allow each transit to have an
independent value of $T_c$. Thus, the period $P$ is relevant to the
model only through the connection between the total mass and the
orbital semi-major axis, $a$. We fix $P=2.51997$~days for WASP-1 and
$P=2.152226$~days for WASP-2, as determined by Collier Cameron et
al.~(2006). The uncertainties in $P$ are negligible for our purposes.

The values of $R_{\star}$ and $R_p$ that are inferred from the
photometry are covariant with the stellar mass.  For a fixed period
$P$, the characteristics of the transit light curve depend almost
exactly on the combinations $R_{\star}/M_{\star}^{1/3}$ and
$R_p/M_{\star}^{1/3}$. Our approach was to fix $M_{\star}$ at the
value reported by Collier Cameron et al.~(2006), which they derived by
comparing the spectroscopically-estimated effective temperatures and
surface gravities to theoretical evolutionary tracks for stars of
different masses.  We then used the scaling relations for the fitted
radii, $R_p \propto M_{\star}^{1/3}$ and $R_{\star} \propto
M_{\star}^{1/3}$, to estimate the systematic error due to the
uncertainty in $M_{\star}$.

To calculate the relative flux as a function of the projected
separation of the planet and the star, we assumed the limb-darkening
law to be quadratic,
\begin{equation}
\frac{I(\mu)}{I(1)} = 1 - u_1(1-\mu) - u_2(1-\mu)^2,
\end{equation}
where $I$ is the intensity, and $\mu$ is the cosine of the angle
between the line of sight and the normal to the stellar surface.  We
employed the analytic formulas of Mandel \& Agol~(2002) to compute the
integral of the intensity over the unobscured portion of the stellar
disk. We fixed the limb-darkening parameters $u_1$ and $u_2$ at the
values calculated by Claret~(2004) for a star with the
spectroscopically-estimated effective temperature and surface
gravity. For WASP-1, these values are $u_1=0.1517$, $u_2=0.3530$; for
WASP-2, they are $u_1=0.2835$, $u_2=0.2887$. We also investigated the
effects of changing the limb-darkening law and allowing the
limb-darkening parameters to vary in the fit (see below).

Each transit also requires two additional parameters for its
description: the out-of-transit flux $f_{\rm oot}$, and a residual
extinction coefficient $k$. The latter is defined such that the
observed flux is proportional to $\exp(-kz)$ where $z$ is the
airmass. In total, there are 6 adjustable parameters for each transit:
$R_{\star}$, $R_p$, $i$, $T_c$, $f_{\rm oot}$ and $k$.

Our goodness-of-fit parameter is
\begin{equation}
\chi^2 = \sum_{j=1}^{N}
\left[
\frac{f_j({\mathrm{obs}}) - f_j({\mathrm{calc}})}{\sigma_j}
\right]^2 
\end{equation}
where $f_j$(obs) is the flux observed at time $j$, $\sigma_j$ is the
corresponding uncertainty, and $f_j$(calc) is the predicted model
value.  The WASP-1 data set has $N=657$ points (after excluding points
at the beginning of the sequence, as described in \S~2), and the
WASP-2 data set has $N=426$ data points.  As noted in \S~3, we took
the uncertainties $\sigma_j$ to be the calculated uncertainties after
multiplication by a factor specific to each night, such that
$\chi^2/N_{\rm DOF} = 1$ when each night's data were fitted
independently.

We began by finding the values of the parameters that minimize
$\chi^2$, using the venerable AMOEBA algorithm~(Press et al.~1992, p.\
408). Then we estimated the {\it a posteriori} joint probability
distribution for the parameter values using a Markov chain Monte Carlo
(MCMC) technique (for a brief introduction, consult appendix A of
Tegmark et al.~2004). In this method, a chain of points in parameter
space is generated from an initial point by iterating a jump function,
which in our case was the addition of a Gaussian random number to each
parameter value. If the new point has a lower $\chi^2$ than the
previous point, the jump is executed; if not, the jump is only
executed with probability $\exp(-\Delta\chi^2/2)$. We set the typical
sizes of the random perturbations such that $\approx$25\% of jumps are
executed. We created 10 independent chains with 500,000 points each,
starting from random initial positions.  The first 100,000 points were
not used, to minimize the effect of the initial condition.  The Gelman
\& Rubin~(1992) $R$ statistic was close to unity for each parameter, a
sign of good mixing and convergence.

\section{Results}

The model that minimizes $\chi^2$ is plotted as a solid line in
Fig.~\ref{fig:lc}. The optimized residual extinction correction has
been applied to the data that are plotted in Fig.~\ref{fig:lc}, and to
the data that are given in Table~1. The differences between the
observed fluxes and the model fluxes are also shown beneath each light
curve.

Tables~3 and 4 give the estimated values and uncertainties for each
parameter based on the MCMC analysis. They also include some useful
derived quantities: the impact parameter $b= a \cos i / R_{\star}$;
the transit duration (i.e. the elapsed time between first contact
$t_{\rm I}$ and last contact $t_{\rm IV}$); and the duration of
ingress (i.e. the elapsed time between $t_{\rm I}$ and second contact
$t_{\rm II}$).  Although the MCMC distributions are not exactly
symmetric about the median, Tables~3 and 4 report (with two
exceptions) only the median values for the derived parameters and
their standard deviations.  The exceptions are the impact parameter
$b$ and inclination $i$ for WASP-1. Those results are best described
as one-sided confidence limits because the data are consistent with a
central transit.

There are several sources of systematic error that are not taken into
account by the MCMC analysis. The first is the systematic error that
results from the covariance between $M_{\star}$ and both of the
parameters $R_p$ and $R_{\star}$, as discussed in \S4.  For WASP-1, we
adopted $M_{\star}=1.15$~$M_\odot$ based on the interpretation by
Collier Cameron et al.~(2006) of the stellar spectrum. Those authors
report an uncertainty of about 15\% in $M_{\star}$, which translates
into a systematic error of 5\% in our estimates of $R_{\star}$ and
$R_p$. For WASP-2, we adopted $M_{\star}=0.79\, M_{\odot}$, and the
uncertainty in $M_{\star}$ is about 12\%, which in turn contributes a
4\% error in $R_{\star}$ and $R_p$.\footnote{We note that our formal
systematic errors should be asymmetric because Collier Cameron et al.\
2006 reported asymmetric error bars on $M_{\star}$, which we have not
taken into account here.}  The other transit parameters (such as $b$,
$i$, and $T_c$) do not depend on $M_{\star}$.

A second potential source of systematic error is the bias due to an
incorrect choice of either the limb-darkening function or the values
of the limb-darkening coefficients.  We investigated the effects of
varying the functional form of the limb-darkening law from quadratic
to linear, and of allowing the coefficients to be free parameters
rather than holding them fixed, and in all cases we found that the
resulting changes to $R_p$ were much smaller than the other sources of
error. We conclude that the systematic error in $R_p$ due to the
choice of limb-darkening law is small compared to either the
statistical uncertainty or the systematic uncertainty due to the
covariance with $M_{\star}$.

\section{Discussion}

\begin{figure}[p]
\epsscale{1.0}
\plotone{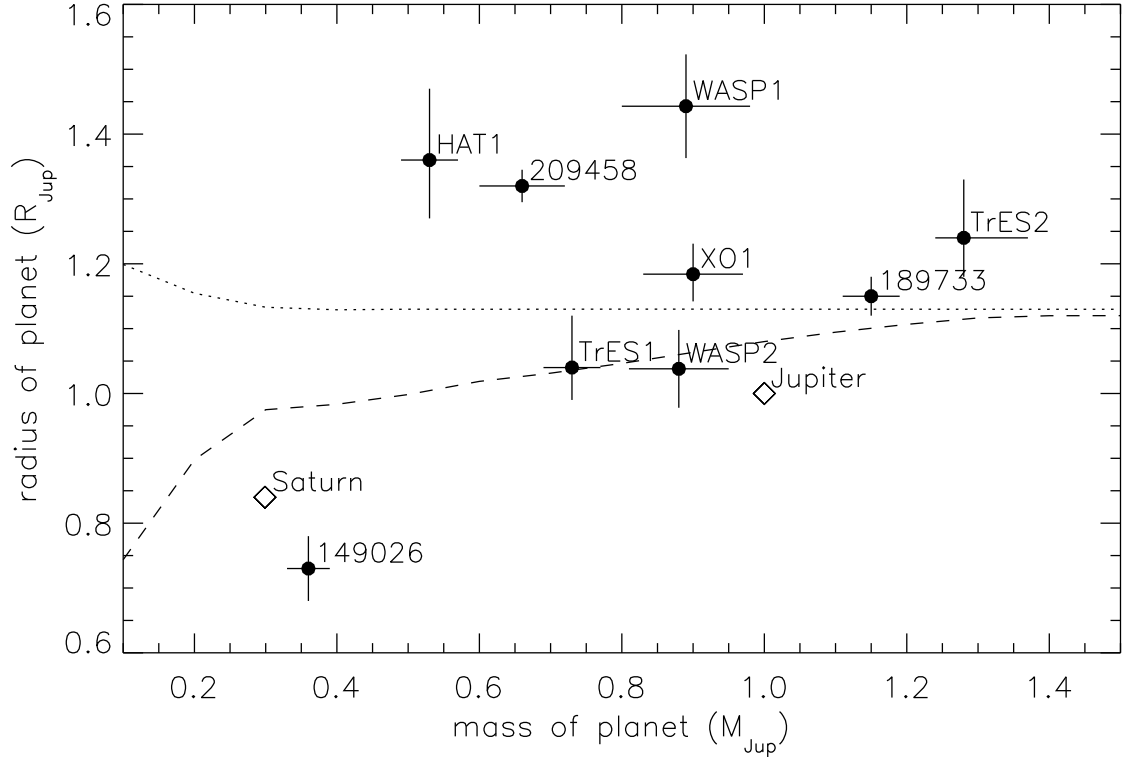}
\caption{
Masses and radii for the known transiting extrasolar planets within 300~pc,
as well as Jupiter and Saturn for comparison.
The dotted line corresponds to the insolated coreless structural models of 
Bodenheimer et al.\ (2003) for an age of 4.5~Gyr and a planetary 
effective temperature of 1500~K.  The dashed line shows their models
for the same parameters but including the presence of a 20-$M_{\Earth}$ core
of solid material.  
Insolation alone is clearly insufficient to account for the large radii
of three of the planets (HAT-P-1b, WASP-1b, and HD~209458b) and likely
a fourth (TrES-2), regardless of whether or not a core is present.
Interestingly, the parent stars of these four planets are significantly
more massive than those of the planets that are in good agreement with
the models (TrES-1, WASP-2b, and HD 189733b), all of which orbit lower-mass,
K dwarf stars.
\label{fig:mr}}
\end{figure}

Our revised estimates for $R_p$ for both WASP-1b and WASP-2b are five
times more precise than those presented in the discovery paper.  The
three exoplanets WASP-2b, XO-1b (McCullough et al.\ 2006; Holman et
al.\ 2006a), and WASP-1b present an interesting sequence (Fig.~2):
their radii differ by as much as 40\%, despite their indistinguishable
masses.  We note that the radius of WASP-2b is in good agreement with
published structural models that include both a 20~$M_{\Earth}$ core
of solid material and the effects of stellar insolation (Bodenheimer
et al.\ 2003).  The radius of XO-1b is larger, but it can be explained
by a coreless model of a similar effective temperature (Fig.~2).  In
contrast, we find that WASP-1b is significantly larger than such
predictions, whether or not a core is included.  WASP-1b is not alone
in its anomalous size: both HD~209458b (Knutson et al.\ 2006) and
HAT-P-1b (Bakos et al.\ 2006a) also require an additional source of
internal energy to account for their large radii.  We also note that
TrES-2 (O'Donovan et al.\ 2006) may require such heating as well,
depending on the outcome of more precise measurements of the planetary
radius.

Only a month ago, HD~209458b was the single known case of a hot
Jupiter that is almost certainly too large to be explained by standard
models of planetary structure. (The other possible case, OGLE-TR-10,
was ambiguous because of the uncertainty in its radius; see Holman et
al.~2006b.) With only one strong anomaly, explanations requiring
somewhat improbable events were perfectly viable.  However, now that a
significant fraction of the transiting hot Jupiters are found to be
similarly in need of this additional energy, the burden of the
theorists may shift to seeking explanations for this effect that are
more generally applicable.

Examining Fig.~2, we note that the three planets in closest agreement
with the published structural models of Bodenheimer et al. (2003) all
orbit the lowest-mass stars of the sample, namely TrES-1 (Alonso et
al.\ 2004; Sozzetti et al.\ 2004; Laughlin et al.\ 2005b), 
WASP-2, and HD~189733 (Bouchy et al.\ 2005; Bakos et al.\ 2006b), 
whereas the primary stars of the
three largest hot Jupiters all orbit stars more massive than the Sun.
Although it is likely too soon to search for such patterns in these
data (we note that the planet of the most massive star, HD~149026, is
the smallest of the sample), we are encouraged that the recent rapid
rate of detection of transiting hot Jupiters will soon provide us with
a signficantly larger sample in which to assess this and other
possible correlations.

Another interesting implication of our measurements for both WASP-1
and WASP-2 is that they are both particularly favorable targets for
efforts to detect reflected light from exoplanets. A positive
detection of reflected light would lead to the first empirical
determination of an exoplanetary albedo, and perhaps even its phase
function. However, the reflected light is typically a minuscule
fraction of the direct light from the star, which explains the long
list of unsuccessful attempts to measure this signal both in
ground-based spectra (Charbonneau et al.\ 1999; Collier Cameron et
al.\ 2002; Leigh et al.\ 2003a, 2003b) and space-based photometry
(Rowe et al.\ 2006). Since the points of first and last contact 
correspond to orbital phase angles that are 
typically within 10$^{\circ}$ of opposition, we may estimate the ratio
of the planetary flux $f_p$ to that of the star $f_{\star}$ to be $f_p
/ f_{\star} \simeq p\, (R_{p} / a)^{2}$, where $p$ denotes the
wavelength-dependent geometric albedo. For WASP-1, this quantity is $p
\times 3.3 \times 10^{-4}$, the most favorable for any known
transiting system.  The other systems for which favorable
planet-to-star contrast ratios are expected are HD~189733 ($p \times
3.1 \times 10^{-4}$), TrES-2 ($p \times 2.8 \times 10^{-4}$), and
WASP-2 ($p \times 2.7 \times 10^{-4}$).  The contrast ratios for all
of these systems are superior to those for the systems that have been
studied to date.  We note that the long duration of the WASP-1 transit
(the consequence of a nearly equatorial transit of a large star)
further facilities a search for reflected light, as it increases the
total time in which to gather the signal. Binning the data for WASP-1
in Fig.~1 would yield, in principle, a photon-noise limited precision
of $9.5 \times 10^{-5}$, which is sufficient to address large values
of $p$ with good statistical significance, should we succeed in
obtaining a time series of similar quality spanning a secondary
eclipse.

\acknowledgments We thank Greg Laughlin for providing the theoretical
mass-radius curves shown in Fig.~2.  This material is based upon work
supported by NASA from the {\it Kepler} mission and under grant
NNG05GJ29G issued through the Origins of Solar Systems Program.

\begin{deluxetable}{lcccc}
\tabletypesize{\normalsize}
\tablecaption{Photometry of WASP-1\label{tbl:phot-wasp1}}
\tablewidth{0pt}

\tablehead{
\colhead{HJD} & \colhead{Relative flux} & \colhead{Uncertainty}
}

\startdata
$2454005.64040$ & $1.00235$ & $0.00204$ \\
$2454005.64088$ & $0.99903$ & $0.00202$ \\
$2454005.64138$ & $0.99851$ & $0.00202$ \\
\enddata 

\tablecomments{The time stamps represent the Heliocentric Julian Date
  at the time of mid-exposure. The data have been corrected for
  residual extinction effects, and the uncertainties have been
  rescaled as described in \S3. We intend for this table to appear in
  entirety in the electronic version of the journal. A portion is
  shown here to illustrate its format. The data are also available
  from the authors upon request.}

\end{deluxetable}

\begin{deluxetable}{lcccc}
\tabletypesize{\normalsize}
\tablecaption{Photometry of WASP-2\label{tbl:phot-wasp2}}
\tablewidth{0pt}

\tablehead{
\colhead{HJD} & \colhead{Relative flux} & \colhead{Uncertainty}
}

\startdata
$2454008.60531$ & $0.99881$ & $0.00159$ \\
$2454008.60578$ & $1.00044$ & $0.00159$ \\
$2454008.60627$ & $0.99805$ & $0.00159$ \\
\enddata 

\tablecomments{The time stamps represent the Heliocentric Julian Date
  at the time of mid-exposure. The data have been corrected for
  residual extinction effects, and the uncertainties have been
  rescaled as described in \S3. We intend for this table to appear in
  entirety in the electronic version of the journal. A portion is
  shown here to illustrate its format. The data are also available
  from the authors upon request.}

\end{deluxetable}

\begin{deluxetable}{ccc}
\tabletypesize{\small}
\tablecaption{System Parameters of WASP-1\label{tbl:params-wasp1}}
\tablewidth{0pt}

\tablehead{
\colhead{Parameter} & \colhead{Value} & \colhead{Uncertainty}
}
\startdata
                                $R_{\star}/R_\odot$& $          1.453$ & $          0.032$ \\
                            $R_p/R_{\rm J}$& $          1.443$ & $          0.039$ \\
                                  $R_p / R_{\star}$& $        0.10189$ & $        0.00093$ \\
                                    $i$~[deg]& $     > 86\fdg 1$ &    (95\% conf.) \\
                                          $b$& $        < 0.336$ &    (95\% conf.) \\
                $t_{\rm IV} - t_{\rm I}$~[hr]& $          3.773$ & $          0.031$ \\
               $t_{\rm II} - t_{\rm I}$~[min]& $           21.5$ & $            1.1$ \\
                                  $T_c$~[HJD]& $  2454005.75196$ & $        0.00045$
\enddata

\tablecomments{The parameter values in Column 2 are the median values
  of the MCMC distributions, and the uncertainties in Column 3 are the
  standard deviations. These are for a {\it fixed} choice of
  $M_{\star}=1.15~M_\odot$, and
  for a {\it fixed} choice of the limb-darkening function (see the
  text). The 15\%
  uncertainty in $M_{\star}$ introduces an {\it additional} 5\%
  uncertainty in $R_{\star}$ and $R_p$ (and has no effect on the other
  parameters).}

\end{deluxetable}

\begin{deluxetable}{ccc}
\tabletypesize{\small}
\tablecaption{System Parameters of WASP-2\label{tbl:params-wasp2}}
\tablewidth{0pt}

\tablehead{
\colhead{Parameter} & \colhead{Value} & \colhead{Uncertainty}
}
\startdata
                                $R_{\star}/R_\odot$& $          0.813$ & $          0.032$ \\
                            $R_p/R_{\rm J}$& $          1.038$ & $          0.050$ \\
                                  $R_p / R_{\star}$& $         0.1309$ & $         0.0015$ \\
                                    $i$~[deg]& $          84.74$ & $           0.39$ \\
                                          $b$& $          0.731$ & $          0.026$ \\
                $t_{\rm IV} - t_{\rm I}$~[hr]& $          1.799$ & $          0.035$ \\
               $t_{\rm II} - t_{\rm I}$~[min]& $           24.6$ & $            2.4$ \\
                                  $T_c$~[HJD]& $  2454008.73205$ & $        0.00028$
\enddata

\tablecomments{The parameter values in Column 2 are the median values
  of the MCMC distributions, and the uncertainties in Column 3 are the
  standard deviations. These are for a {\it fixed} choice of
  $M_{\star}=0.79~M_\odot$, and
  for a {\it fixed} choice of the limb-darkening function (see the
  text). The 12\%
  uncertainty in $M_{\star}$ introduces an {\it additional} 4\%
  uncertainty in $R_{\star}$ and $R_p$ (and has no effect on the other
  parameters).}

\end{deluxetable}

\end{document}